\documentclass[12pt]{article}
\usepackage{latexsym}
\newcommand{\call}{{\cal L}}
\newcommand{\calm}{{\cal M}}
\newcommand{\calf}{{\cal F}}

\newcommand{\calr}{{\cal R}}
\newcommand{\calj}{{\cal J}}

\newcommand{\llp}{\ell(\ell+1)}

\newcommand{\kma}{\; ,}
\newcommand{\pkt}{\; .}
\newcommand{\be}{\begin{equation}}
\newcommand{\ee}{\end{equation}}
\newcommand{\bea}{\begin{eqnarray}}
\newcommand{\eea}{\end{eqnarray}}
\newcommand{\bfx}{{\bf x}}
\newcommand{\bfm}{{\bf M}}
\newcommand{\bfphi}{{\bf \Phi}}
\newcommand{\phidagphi}{\bfphi^\dagger\bfphi}

\begin{document}
\begin{titlepage}
\begin{flushright}
DO-TH-99/08\\
May 1999
\end{flushright}
\vspace{20mm}
\begin{center}
{\Large \bf
Gauge invariance of the one-loop effective action
of the Higgs field in the $SU(2)$ Higgs model}

\vspace{10mm}
{\large  J\"urgen Baacke\footnote{
 e-mail:~baacke@physik.uni-dortmund.de} and Katrin Heitmann
\footnote{e-mail:~heitmann@hal1.physik.uni-dortmund.de}
\vspace{15mm}}

{\large Institut f\"ur Physik, Universit\"at Dortmund} \\
{\large D - 44221 Dortmund , Germany}
\vspace{10mm}

{\bf Abstract}
\end{center}

The one-loop effective action of the abelian and nonabelian
Higgs models has been studied in various gauges, in the context of
instanton and sphaleron transition, bubble nucleation and most recently
in nonequilibrium dynamics. Gauge invariance is expected on account of
Nielsen' s theorem, if the classical background field is an extremum
of the classical action, i.e., a solution of the classical equation of motion. 
We substantiate this general statement for the one-loop effective
action, as computed using mode functions. We show that in the gauge-Higgs
sector there are two types of modes that satisfy the same equation of motion
as the Faddeev-Popov modes. We apply the general
analysis to the computation of the fluctuation determinant
for bubble nucleation in the $SU(2)$ Higgs model in the
't Hooft gauge with general gauge parameter $\xi$.
We show that due to the cancellation of the modes mentioned above the
fluctuation determinant is independent of $\xi$. 

\end{titlepage}
\section{Introduction}
The effective potential of gauge theories has been
considered extensively as it is of interest for 
the phase structure of these theories, and in
particular for the discussion of phase transitions 
and the associated bubble nucleation rates
\cite{Enq:1992,Baa:1993b,Baa:1995a,Kripf:1995,Baa:1996,Sur:1998}. 
More generally the effective action appears when
computing fluctuation corrections to the sphaleron 
\cite{Carson:1990a,Carson:1990b,Baacke:1994a,Diakonov:1994} 
and instanton \cite{Baacke:1995b}
transition rates in such theories. 

It is well known that the effective potential is gauge dependent
except for the region around the extrema of the effective action,
where Nielsen' s theorem states that the gauge dependence 
should disappear \cite{Nielsen:1975}. 
This has been verified in various cases 
\cite{Aitchison:1984,Kobes:1991,Contreras:1997}. 
Besides the static extrema such as the minima and maxima
of the effective {\em potential} Nielsen' s theorem 
more generally applies to
the  extrema of the effective {\em action}, i.e. to the extremal, classical
 paths in configuration space, such as the 
bubble and sphaleron actions.
It has been verified  \cite{Metaxas:1996}, using the 
gradient expansion, that for the leading orders
in the coupling the quantum corrections to the bubble nucleation rate are 
gauge independent. Exact numerical computations are based on
the analysis and numerical computation of mode functions in the
 background of the classical solution. Such computations are 
quite demanding numerically, as well as algebraically and analytically,
so in general the authors just used one particular gauge,
as e.g. the 't Hooft Feynman gauge and a concise discussion of
gauge independence is lacking.

We have recently analyzed the evolution equations for a Higgs condensate
and the gauge and Higgs field fluctuations in the $SU(2)$ Higgs model,
in one-loop approximation \cite{Baacke:1997b}. 
Here again the question of gauge 
dependence arises and has not yet been analyzed. For the abelian
Higgs model, a gauge invariant formalism has been developed, which, 
however, has not yet been implemented numerically 
\cite{Boyanovsky:1996,Boyanovsky:1998}.

We consider here the $SU(2)$ Higgs model with an isoscalar
Higgs background field. Such a configuration plays a 
central r\^ole in the discussion of the electroweak phase
transition. Its finite temperature effective potential has been discussed 
extensively, and it has been used to predict the
rate of bubble nucleation in a first order phase transition 
\cite{Enq:1992,Baa:1993b,Baa:1995a,Kripf:1995,Baa:1996,Sur:1998}.

The plan of the paper is as follows: in the section 2 we
present the basic equations, and  we expand the Lagrangian into a
classical and a second order fluctuation part. In section 3 we
 present the equations of motion without gauge-fixing, and in
the 't Hooft background gauge for arbitrary gauge parameter $\xi$.
 We show that there are two types of modes, the
gauge modes and the gauge-fixing modes that satisfy the 
same equations of motion as the Faddeev-Popov ghosts,
{\em if the classical background field satisfies its
equation of motion}. This 
observation is the clue for a reduction of the mode equations
into equations for the physical degrees of freedom and
into equations whose functional determinant is cancelled
by the Faddeev-Popov one. This reduction depends on the system
under consideration. Here we demonstrate the cancellation 
of the unphysical modes for the fluctuation determinant
which determines the fluctuation corrections to
bubble nucleation, in the $SU(2)$ Higgs model.
We briefly introduce the model and its fluctuation operator
in section 4. The coupled gauge-Higgs system is analyzed
in the partial-wave reduced equations in section 5.
After some suitable transformations the system is reduced
to a triangular form, with the consequence that the 
fluctuation determinant can be computed from the
diagonal part. Thereby the cancellation against the
Faddeev-Popov contributions to the fluctuation
determinant is demonstrated explicitly.
We present some conclusions in section 6.


\section{Fluctuation Lagrangian and mode equations}
\setcounter{equation}{0}
The Lagrangian of the $SU(2)$ Higgs model reads
\begin{equation}
{\cal L}=-\frac{1}{4}F_{\mu\nu}^aF^{a\mu\nu}+\frac{1}{2}(D_\mu\Phi)^\dagger
(D^\mu\Phi)-V(\Phi^\dagger\Phi)\kma
\end{equation}
with the field strength tensor
\begin{equation}
F_{\mu\nu}^a
=\partial_\mu{A_\nu^a}-\partial_\nu{A_\mu^a}+g\epsilon^{abc}
{A_\mu^b}{A_\nu^c}\kma
\end{equation}
and the covariant derivative
\begin{equation}
D_\mu \equiv \partial_\mu-i \frac{g}{2} A_\mu^a\tau^a\pkt
\end{equation}
The potential has the form
\begin{equation}
V(\Phi^\dagger\Phi)=\frac{\lambda}{4}(\Phi^\dagger\Phi-v^2)^2\pkt
\end{equation}
We will assume in the following a classical field (condensate)
\be
\Phi(x)=H(x)\left(\begin{array}{c}0 \\ 1\end{array}
\right)
\kma\ee
its space-time dependence is not further specified here.
A time - independent, metastable, radially symmetric configuration
will be relevant for bubble nucleation, a spatially homogenous
time dependent field describes a nonequilibrium situation,
as considered in \cite{Baacke:1997b}.
The fluctuations around this space-time dependent condensate are 
parameterized as
\be
\Phi(x)=[H(x)+h(x)+i\tau_a\phi_a(x)]
\left(\begin{array}{c}0\\1\end{array}\right)
\kma\ee
with the isoscalar Higgs mode $h(x)$ and the would-be Goldstone
fields $\phi_a(x)$, $a=1\dots 3$.
As there is no classical gauge field, we have
\be
A_a^\mu(x)=a_a^\mu(x) \pkt
\ee
The Lagrangian can then be split into a classical part
\be
\call_{\rm cl}(x)=\frac{1}{2}\left[\partial_\mu H\partial^\mu H
-\frac{\lambda}{4}(H^2-v^2)^2\right]
\ee
 and a fluctuation Lagrangian.  
 The part of first
order in the fluctuating field vanishes, if the classical equation of motion
\be \label{Hem}
\Box H +\lambda (H^2-v^2)H =0
\ee
is fulfilled.
The part of second order in the fluctuations reads
\bea
\label{L2nd}
\call^{(2)} &=& \frac{1}{2}\left\{
-\partial_\mu a^a_\nu \partial^\mu
a_a^\nu + \partial_\mu a^a_\nu \partial^\nu a_a^\mu
\right. \nonumber\\
&&  +\frac{g^2}{4} H^2 a^a_\mu a_a^\mu
+\partial_\mu \phi_a \partial^\mu \phi_a
+g \partial_\mu H a^\mu_a \phi_a -
g  H a_\mu \partial^\mu\phi_a
\\ \nonumber &&\left. -\lambda\left (H^2-v^2\right)\phi_a\phi_a
+\partial_\mu h \partial^\mu h
-\lambda(3 H^2-v^2) h^2 \right\} \pkt
\eea
In the one-loop approximation we do not have to consider
higher order terms.

The gauge-fixing term, in the 't Hooft background gauge is given by
\be
\call^{(2)}_{\rm gf} = -\frac{1}{2\xi}\calf_a\calf_a
\ee
with the gauge conditions
\be
\calf_a=\partial_\mu a_a^\mu +\xi e H \phi_a
\kma\ee
the Faddeev-Popov Lagrangian is
\be
\call_{\rm FP}=\frac{1}{2}
\left\{ \partial_\mu\eta_a \partial^\mu \eta_a
-\xi \frac{g^2}{4} H^2 \eta_a\eta_a \right \}
\pkt
\ee


\section{Gauge mode and gauge-fixing mode}
\setcounter{equation}{0}

Before discussing the fluctuation operator for a specific
physical setting we specify here the unphysical degrees of 
freedom in the gauge field
and would-be Goldstone sector whose cancellation against 
the Faddeev-Popov modes will lead to a gauge invariant fluctuation
determinant. The fluctuation operator of the isoscalar Higgs mode $h(x)$
is gauge invariant from the outset.

We arrange the gauge field fluctuations
$a_a^\mu$  and
the would-be Goldstone fields $\varphi_a$ in a 
$(4+1)$ column vector
\be
\psi_a=\left\{\begin{array}{c}a_a^\mu\\\phi_a\end{array}\right\}\pkt
\ee
We start with the equations of motion obtained 
{\em without} the gauge-fixing term.
The differential operator (fluctuation operator) 
governing the mode evolution then
takes the form
\be
\calm =\left\{\begin{array}{cc}
- (\Box + \frac{g^2}{4} H^2) \delta_\mu^\nu +\partial^\nu
\partial_\mu
& - \frac{g}{2}\partial^\nu H +\frac{g}{2}H\partial^\nu\\
-g  \partial_\mu H  -\frac{g}{2} H \partial_\mu  & \Box + \lambda (H^2-v^2)
\end{array}\right\}
\pkt\ee
The mode equations are  the same for all $a=1,2,3$ ,
\be \label{meuf}
\calm \psi_a =0 \pkt
\ee
An infinitesimal gauge transform  is given by
\be
\psi_a^g(x) =\left\{\begin{array}{c}
\partial^\mu\\ \frac{g}{2}H(x)\end{array}\right\} f_a(x)
\pkt\ee
These modes satisfy the mode equation (\ref{meuf})
 if $H(x)$ satisfies
the classical field equation (\ref{Hem}). The latter condition
is crucial. It arises from the mode equation for $\varphi_a$, 
the one for the
vector potentials is fulfilled trivially.

If the gauge mode is substituted into the gauge condition
one finds
\be
\left( \calf_a \right)_{\rm g} = \left[\Box  + 
\xi \frac{g^2}{4} H^2(x)\right] f_a 
\kma\ee
the differential operator on the r.h.s. is just the
Faddeev-Popov operator. So, if the gauge mode is
inserted into the Lagrangian, the gauge-fixing term contains
the Faddeev-Popov operator {\em squared}. It is very suggestive that
the contribution of this squared operator to the effective action
, i.e., to the $\log \det $ of the fluctuation operator,
is cancelled by twice the $\log \det$ of the Faddeev-Popov
operator. 

If the gauge-fixing term is included the
fluctuation operator takes the form 
\be
\calm_{\rm f} =\left\{\begin{array}{cc}
-(\Box+\frac{g^2}{4} H^2) \delta_\mu^\nu +
\left(1-\frac{1}{\xi}\right)\partial^\nu\partial_\mu
&-g\partial^\nu H \\
-g \partial_\mu H   &\Box + \lambda (H^2-v^2)
+\frac{g^2}{4}\xi H^2
\end{array}\right\}
\pkt\ee
If we apply the fluctuation operator to the gauge mode, and use the classical
equation of motion, we obtain
\be
\calm_{\rm f}\psi^g_a(x)=\left\{\begin{array}{c}
-\frac{1}{\xi}\partial_\mu \\ \frac{g}{2} H(x)\end{array}\right\}
\left[\Box+\xi \frac{g^2}{4} H^2(x)\right] f_a(x) =\calm_{\rm FP} f_a(x)
\pkt
\ee
The differential operator appearing on the right hand side is just
the Faddeev-Popov operator
\be \label{FPfluct}
\calm_{\rm FP} =\Box+\xi \frac{g^2}{4} H^2(x)
\pkt
\ee
If $f_a$ is an eigenfunction of the Faddeev-Popov operator,
$\calm_{\rm FP}f_a=\omega^2_{\rm FP} f_a$, then the associated gauge mode
satisfies
\be
\left\{\begin{array}{cc}
-\xi & 0\\ 0 & 1\end{array}\right\}\calm_{\rm f}\psi^g_a=
\omega^2_{\rm FP}\psi^g_a
\pkt\ee
The factor $\xi$ in the matrix multiplies the four gauge field 
components. So the fluctuation operator modified by multiplication
with a constant matrix, has a class of eigenfunctions with the
same eigenvalues as the Faddeev-Popov operator. In the 
effective action the modification by the constant matrix
is irrelevant, as one computes the ratio
between  the fluctuation determinants in the background field 
and in a standard vacuum configuration, to which the same arguments apply.

Now consider the gauge condition  $\calf_a$. We introduce the covector
\be
{\bf u_\xi} = \left[\partial_\mu,\xi \frac{g}{2} H(x)\right]\kma
\ee
so that
\be
\calf_a={\bf u_\xi}\phi_a
\pkt\ee
Consider an arbitrary mode $\psi_a$.
We then find, using again the classical equation of motion,
\be
{\bf u_\xi}\left\{\begin{array}{cc}
-\xi & 0\\ 0 & 1\end{array}\right\}\calm_{\rm f}\psi_a
=\left[\Box  +\xi \frac{g^2}{2} H^2(x)\right]{\bf u_\xi}\psi_a
=\calm_{\rm FP}\calf_a
\pkt
\ee
Let $\psi^\alpha_a$ now be an eigenmode of the modified fluctuation operator
with eigenvalue $\omega^2_\alpha$.
Then this equation entails
\be
{\bf u_\xi}\omega^2_\alpha\psi^\alpha_a=\left(\omega^\alpha_a\right)^2
\calf_a^\alpha
=\calm_{\rm FP}\calf_a^\alpha
\pkt
\ee
So if the projection on the vector ${\bf u_\xi}$ is different from zero
the eigenvalue is simultaneously an eigenvalue of $\calm_{\rm FP}$.
We thereby have a second class of modes on which the 
fluctuation operator of the gauge-Higgs system has
the same spectrum as the Faddeev-Popov operator.
We call them gauge-fixing modes. We have to make sure
that this class of modes, obtained by a projection, is not empty, and not
identical with the gauge modes. 

Obviously the modes on which 
the projector ${\bf u_\xi}$ yields zero
are those which satisfy the gauge condition, these are the ``physical
modes''. We know that out of the five components of the
gauge-Higgs modes $\psi$ only three are physical, they represent
the spatial components of the massive gauge field.

We next consider the action of the projector on the
gauge eigenmodes.
It is convenient to introduce a vector ${\bf v}$ that generates the
gauge modes via
\be
\psi^g_a={\bf v}f_a =\left\{\begin{array}{c}
\partial^\mu \\ \displaystyle \frac{g}{2}H(x)\end{array}\right\} f_a
\pkt
\ee
We  note that
\be
{\bf u_\xi}{\bf v}=\Box+\xi \frac{g^2}{4} H^2
\pkt
\ee
This implies that the gauge-fixing mode obtained by projection
of a gauge mode
satisfies
\be \label{gaugeproj}
\calf_a={\bf u_\xi}\psi^g_a={\bf u_\xi}{\bf v}f_a=
\left[\Box+\xi\frac{g^2}{4} H^2(x)\right]
f_a
\pkt\ee
So if $f_a$ is an eigenfunction of the Faddeev-Popov operator, 
then the gauge-fixing mode generated from it does not represent
a new, independent mode. 
However, the gauge modes and the physical modes do not exhaust
the Hilbert space that is based on five field degrees of freedom, and we are
sure that the projector does not give zero on the remaining
subspace.

We have shown up to now, that for a background field satisfying
the classical equation of motion there are two classes of modes
whose contribution to the effective action will be cancelled by the
one of the Faddeev-Popov sector. We have not shown, thereby, that 
the remaining ``physical'' part of the gauge-Higgs sector becomes
independent of $\xi$. Furthermore, the way in which the modes are
eliminated is a technical matter, it
depends on the structure of the background field, and
on the problem under consideration. So if we want to
illustrate the application of these general results we have to 
consider specific models.

We will here analyze the modes introduced above, and the 
cancellation of their
contribution to the fluctuation determinant, for the case of
bubble nucleation in the $SU(2)$ Higgs model.


\section{Bubble nucleation in the $SU(2)$ Higgs model}
\setcounter{equation}{0}
Bubble nucleation occurs in the $SU(2)$ Higgs model if the
phase transition from the symmetric high temperature phase
to the broken symmetry phase at low temperature is first order.
It has been considered as providing a possible mechanism for
baryogenesis, a possibility ruled out by the present lower limit
for the Higgs mass. Still the model is of interest, in particular
it can be studied in lattice simulations for sufficiently low Higgs 
masses. The phase transition is described (see, e.g., \cite{DLHLL}), 
by the 3-dimensional high-temperature action
\bea \label{htac}
S_{ht} & =&\frac{1}{g_3(T)^2} 
\int d^3x \left[\frac{1}{4}F_{ij}F_{ij}+
\frac{1}{2}(D_i\Phi)^\dagger (D_i \Phi)
 + V_{ht}(\phidagphi) \right. \nonumber \\
&&\left.+ \frac{1}{2} A_0 \left(-D_iD_i +\frac{1}{4} \phidagphi
\right )A_0 \right] \; .
\eea
Here the coordinates and fields have been rescaled as \cite{Carson:1990b}
\be \label{scaling1}
\vec x \to \frac{\vec x}{g v(T)}, \;
\bfphi \to v(T) \bfphi, \; A \to v(T) A \; .
\ee
The vacuum expectation value  $v(T)$ is defined as
\be \label{scale1}
v^2(T)=\frac{2D}{\lambda_T} (T_0^2 - T^2)\;.
\ee
$T_0$ is the temperature at which the high-temperature
potential $V_{ht}$ changes its extremum at 
$\bfphi = 0$ from a minimum at $T > T_0$ to a maximum at
$T < T_0$. The temperature dependent coupling of the
three-dimensional theory is defined as
\be \label{htcoupl}
g_3^2(T) = \frac{gT}{v(T)} \; .
\ee
We use the standard parameters
\bea \label{coeffs}
D&=& (3m_W^2+2m_t^2)/8v_0^2\kma  \\
E&=& 3 g^3/32 \pi \kma \\
B&=& 3 ( 3m_W^4 - 4 m_t^4)/64\pi^2v_0^4 \kma \\
T_0^2&=& (m_H^4-8 v_0^2 B)/4D  \kma \\
\lambda_T&=& \lambda -3\left(3m_W^4\ln\frac{m_w^2}{a_BT^2}
-4 m_t^4 \ln\frac{m_t^2}{a_F T^2}\right)/16 \pi^2 v_0^4 \; .
\eea
We use in the following a somewhat different rescaling, introduced
in \cite{HeKriSch,KriLaSch}, based on the secondary minimum 
of the high-temperature potential which occurs at
\be
\tilde v (T) = \frac{3 E T}{2\lambda}+
\sqrt{\left(\frac{3ET}{2\lambda}\right)^2+v^2(T)} \; .
\ee
The high-temperature potential then takes the form 
\be \label{htpot2}
V_{ht}(\phidagphi) =
\frac{\lambda_T}{4g^2} \left\{ (\phidagphi)^2-
\epsilon(T) (\phidagphi)^{3/2} + \left[\frac{3}{2}
\epsilon(T)-2\right]\phidagphi \right\}
\ee
with
\be \label{epsdef}
\epsilon(T)=\frac{4}{3}\left ( 1 - \frac{v(T)^2}{\tilde v (T)^2} 
\right ) \; .
\ee

The standard formula \cite{La,Co,CaCo,Af,Li} for the bubble 
nucleation rate is given by
 \be \label{rate}
\Gamma/V = \frac{\omega_-}{2 \pi} \left (
\frac{\tilde S}{2\pi}\right )^{3/2}\exp(-\tilde S)~ {\calj}^{-1/2}
\; . \ee
Here $\tilde S$ is the high-temperature action, Eq. (\ref{htac}),
minimized by a classical minimal bubble configuration (see below), 
$\cal J$ is the fluctuation
determinant which describes the next-to-leading part of the
semiclassical approach and which will be defined below; 
its logarithm is related to the 1-loop
effective action by 
\be
S^{1-l}_{eff} = \frac{1}{2} \ln {\cal J} \; .
\ee  
Finally $\omega_-$ is the absolute value of
the unstable mode frequency.

The classical bubble configuration is described by
a vanishing gauge field and a real spherically symmetric
Higgs field 
$H(r) = |\bfphi| (r)$ which is a solution of the
Euler-Lagrange equation   
\be \label{Clbub}
-H''(r)-\frac{2}{r}H'(r)+\frac{d V_{ht}}{dH(r)} = 0
\ee
with the boundary conditions 
\be
\lim_{r\to\infty}
H(r)=0 ~~ 
 {\rm and} ~~ H'(0)=0 \; .
\ee 

We expand the gauge and Higgs fields around this classical
configuration via
\bea
W_\mu^a(\bfx) & = & a_\mu^a(\bfx) \nonumber \kma \\
\bfphi(\bfx) & = & \left[H(r) + h(\bfx) + \tau^a \phi_a(\bfx)\right]
\left( \begin{array}{c}
 0 \\ 1 \end{array} \right) \kma 
\eea
where $ a_\mu^a, h $
 and $ \phi_a$ are the fluctuating fields, denoted collectively
by $\varphi_i$.

If the action is expanded with respect to the fluctuating fields,
the first order term vanishes if $H(r)$ satisfies the
classical equation of motion (\ref{Clbub}). The second order part
defines the fluctuation operator via
\be
S^{(2)} =\frac{1}{\tilde g_3^2(T)} 
\int d^3x \frac{1}{2} \varphi_m \calm_{mn}\varphi_n
\pkt
\ee 
The fluctuation determinant $\cal J$ appearing in the rate formula
is defined by \footnote{We omit some sophistications related to 
zero and unstable modes.}
\be
{\cal J} =\frac{\det{\cal M}}{\det {\cal M}^0} \; ,
\ee
where $\calm_0$ is the fluctuation
operator obtained by expanding around a spatially homogenous
classical field that is a minimum of effective potential. 
The gauge conditions for the 3-dimensional theory read
\be \label{backgauge}
{\cal F}_a = \partial_\mu a^\mu_a+ \frac{\xi}{2}H \phi_a =0 \; .
\ee
The total gauge-fixed action $S_t$ is obtained from the high-temperature
action by adding to it the gauge-fixing action
\be
S_{gf} =\frac{1}{\tilde g_3^2(T)} 
\int d^3x \frac{1}{2\xi} {\cal F}_a {\cal F}_a
\kma \ee 
 the corresponding Faddeev-Popov action reads
\be
S_{FP} = \frac{1}{\tilde g_3^2(T)}
 \int d^3x \eta^\dagger \left(-\Delta + \xi \frac{H^2(r)}{4}\right) \eta \; .
\ee
The fluctuation operator is obtained from
the total action $S_t= S_{ht}+S_{gf}+S_{FP}$.
The fluctuation operator, and along with it the fluctuation determinant,
decomposes under partial wave expansion into fluctuation operators 
for fixed angular momentum. It is these that we will consider in the
following.

The background field is isoscalar, so the isospin index $a$ 
just results in multiplicity factors, we will omit it in the following.
The scalar fields $h(\bfx), \phi_a(\bfx), \eta(\bfx)$,
and $a_0(\bfx)$ are expanded
w.r.t. spherical harmonics $Y_\ell^m(\hat{\bfx})$,
the partial wave mode functions are denoted  by
$f^\ell_h(r), f^\ell_\phi(r),f^\ell_\eta(r)$, and $f^\ell_0(r)$.
The vector spherical harmonics $\hat{\bfx} Y_\ell^m, r \nabla Y_\ell^m$,
and $\vec L Y_\ell^m$ are
used for expanding the space components of the gauge fields via
\be
{\bf a}(\bfx)=\sum_{\ell m}
\left(\frac{f^\ell_a(r)}{\sqrt{\llp}} r \nabla Y_\ell^m+
f^\ell_b(r)\hat{\bfx}Y_\ell^m+\frac{f^\ell_c(r)}{\sqrt{\llp}}
\bfx\times\nabla Y_\ell^m\right)
\pkt\ee
The fluctuation operator is block-diagonal. In the following we consider
just one partial wave and omit the superscript $\ell$. We denote the
partial wave reduction of the fluctuation operator
$\calm$ by ${\bf M^\ell}$, we omit the  superscript, however. 
The components $ f_h(r),f_\eta(r),f_c(r)$, and $f_0(r)$ are decoupled,
the operator has the form
\be
\bfm_{nn} =-\frac{d^2}{dr^2} 
-\frac{2}{r} \frac{d}{dr} + \frac{\llp}{r^2} + m_n^2
+V_{m}(r)
\pkt \ee
The masses are $m_\eta=m_0=m_c=0$ and $m_h=m_H$ with the Higgs mass
\be m_H^2 = \frac{\lambda_T}{4 g^2} (3\epsilon -4) 
\kma \ee
the potentials are
$V_0(r)=V_c(r)=H^2(r)/4$, $V_\eta(r)=\xi H^2(r)/4$ and
\be
 V_h(r) = \frac{\lambda_T}{4 g^2} \left[ 12 H^2(r)-6\epsilon H(r)\right]
\pkt
\ee  
The Faddeev-Popov fluctuations are fermionic and two-fold degenerate, 
as usual.

The modes $f_a, f_b$ and $f_\phi$ are coupled.
The nonvanishing components of the fluctuaion operator are
\bea
{\bf M}_{aa}(r) &=&-\frac{d^2}{dr^2} 
-\frac{2}{r} \frac{d}{dr} + \frac{\llp}{\xi r^2} + \frac{H^2(r)}{4}
\\
{\bf M}_{bb} (r)&=&-\frac{1}{\xi}\left(\frac{d^2}{dr^2} 
+\frac{2}{r} \frac{d}{dr} \right)+ \frac{\llp+2/\xi}{ r^2} 
+ \frac{H^2(r)}{4}
\\
{\bf M}_{\phi\phi}(r) &=&-\frac{d^2}{dr^2} 
-\frac{2}{r} \frac{d}{dr} + \frac{\llp}{ r^2} +
\xi\frac{H^2(r)}{4}  \\ \nonumber 
 && \hspace{10mm} +m_H^2
+\frac{\lambda}{g^2}\left[
H^2(r)-\frac{3}{4}\epsilon H(r)\right]
\\
{\bf M}_{ab}(r)&=&-\frac{\sqrt{\llp}}{\xi r^2}\left[2+(1-\xi)r\frac{d}{dr}
\right]
\\
{\bf M}_{ba}(r)&=&-\frac{\sqrt{\llp}}{\xi r^2}\left[1+\xi-(1-\xi)r\frac{d}{dr}
\right]
\\
{\bf M}_{b\phi}(r)&=&{\bf M}_{\phi b}(r)= - H'(r)
\pkt\eea
The fluctuation operator of this
coupled system is hermitean, as it should, because it arises from
the variation of a Lagrangian. The asymmetry suggested by the
explicit form arises from integrations by parts.
 
The gauge parameter $\xi$ only occurs in the coupled system
and for the Faddeev-Popov modes. The cancellation of the
$\xi$ dependence will have to occur between these two
sectors. They will be analyzed in the next section.


\section{Analysis of the fluctuation operator}
\setcounter{equation}{0}

In analyzing the gauge dependence we will have to consider the
coupled system of the modes $f_a,f_b$, and $f_\phi$, i.e., the radial
mode functions for angular momentum $\ell$.
In analogy to section 3 we consider the fluctuation operator multiplied
from the left by a constant matrix ${\rm diag}(\xi,\xi,1)$.
The eigenvalue problem for the fluctuation operator then takes the form
of the three differential equations for the radial mode functions for 
angular momentum $\ell$:
\bea \label{raddgl1}
-f_a'' -\frac{2}{r} f_a' + \frac{\llp}{\xi r^2} f_a+ 
\frac{H^2(r)}{4}f_a \hspace{10mm} &&
\\ \nonumber  -\frac{\sqrt{\llp}}{\xi r^2}\left[2 f_b + (1-\xi)r
f_b'\right]&=&\frac{\omega^2}{\xi} f_a \kma
\\ \label{raddgl2}
-\frac{1}{\xi}\left(f_b''+\frac{2}{r} f_b' \right)
+ \frac{\llp+2/\xi}{ r^2}f_b + \frac{H^2(r)}{4} f_b
\hspace{10mm} &&
\\ \nonumber
-\frac{\sqrt{\llp}}{\xi r^2}\left[(1+\xi) f_a-(1-\xi)rf_a'\right]
 - H'(r) f_\phi &=& \frac{\omega^2}{\xi} f_b \kma
\\\label{raddgl3}
-f_\phi'' 
-\frac{2}{r} f_\phi' + \frac{\llp}{ r^2} f_\phi +
m_H^2 f_\phi + \xi\frac{H^2(r)}{4} f_\phi \hspace{10mm}&&
\\ \nonumber+
\frac{\lambda}{g^2}\left[H^2(r)-\frac{3}{4}\epsilon H(r)\right]
f_\phi - H'(r) f_b &=& \omega^2f_\phi
\pkt\eea
In view of the general arguments of section 3 we now should 
identify the gauge and the gauge-fixing modes.
A general gauge transformation is parameterized by
a function $\chi(\bfx)$. It can be
expanded into partial waves with respect to spherical harmonics, the
radial mode function is denoted by $f_\chi(r)$.
The gauge mode then takes the form
\bea \nonumber
f^g_a(r)&=&\sqrt{\llp}f_\chi(r)
\\\label{gaugemodl}
f_b^g(r)&=&f_\chi'(r)
\\ \nonumber
f_\phi^g(r)&=&-\frac{H(r)}{2}f_\chi(r)\pkt
\eea
The partial wave amplitude of the gauge-fixing mode $\calf$ is obtained
from the general definition 
\be  \label{fixmodl}
\calf(\bfx)=\nabla {\bf a}(\bfx) +\xi \frac{H(r)}{2}\phi(\bfx)
\pkt\ee
This equation is expanded into partial waves.
The radial mode function of the mode $\calf$ then reads
\be
f_\calf(r)=f_b'(r)+\frac{2}{r} f_b(r)-\frac{\sqrt{\llp}}{r}f_a(r)
+\xi\frac{H(r)}{2}f_\phi(r)
\pkt
\ee
It can be checked, using the basic differential equations
(\ref{raddgl1})-(\ref{raddgl3}) and the differential equation for the
background field (\ref{Clbub}), that the mode $f_\calf$ satisfies 
the differential equation for the Faddeev-Popov modes
\be
- f''_{\rm FP}-\frac{2}{r}  f'_{\rm FP} 
+\frac{\llp}{r^2} f_{\rm FP}
+\xi \frac{H^2(r)}{4} f_{\rm FP}=\omega^2 f_{\rm FP}   
\pkt\ee
Likewise, if the gauge function $f_\chi(r)$ satisfies this 
differential equation then the  mode functions $f^g_n$ generated
from it via  Eq. (\ref{gaugemodl}), satisfy the basic differential equations
(\ref{raddgl1})-(\ref{raddgl3}). This is as to be expected from
the general arguments.
  
We now try to separate the system of differential equations by introducing
a suitable set of new mode functions.
We first eliminate the mode $f_a(r)$ in favor of $f_\calf(r)$,
\be
f_a(r)=- r\frac{f_\calf(r)+f'_b(r)+2 f_b(r)-(\xi/2) H(r)f_\phi(r)}
{\sqrt{\llp}}
\ee
 As mentioned above $f_\calf(r)$ satisfies
\be \label{dglF}
- f''_\calf-\frac{2}{r}  f'_\calf 
+\frac{\llp}{r^2} f_\calf
+\xi \frac{H^2(r)}{4} f_\calf=\omega^2 f_\calf   
\pkt\ee
Having eliminated $f_a(r)$ in this way it cannot be used anymore as gauge mode,
for which now $f_b(r)$ is a possible candidate, however one cannot
use  a simple algebraic substitution. We introduce the new mode function
$f_g(r)$, analogous to $\chi(r)$, and eliminate $f_b(r)$ with the substitution
\be
f_b(r)=\frac{d}{dr}f_g(r)
\pkt\ee
We make the two other amplitudes gauge invariant by defining
\bea
\tilde f_\phi(r)&=&f_\phi(r)+\frac{H(r)}{2}f_g(r)
\\
\tilde f_\calf(r)&=&f_\calf(r)+\omega^2 f_g(r)
\pkt \eea
The latter equation follows from the general relation (\ref{gaugeproj}).
We now have to find the equation of motion for the amplitude 
$f_g(r)$. In view of its close relation to the gauge
function $\chi(r)$ we make the ansatz
 \be \label{dglFb}
- f''_g-\frac{2}{r}  f'_g 
+\frac{\llp}{r^2} f_g
+\xi \frac{H^2(r)}{4} f_g=\omega^2 f_g + \calr(r)   
\pkt\ee
We insert the substitutions into the differential equations
(\ref{raddgl2}), (\ref{raddgl3}) and (\ref{dglF}) for 
the amplitudes $f_b(r), f_\phi(r)$, and $f_\calf(r)$, respectively.
We find, after some algebra, the equation
\bea 
\frac{1}{r}\frac{d}{dr}r^2\calr(r)
&=&\frac{1}{2r}\frac{d}{dr}r^2\left[\xi H(r)\tilde f_\phi(r)- 
 \tilde f_\calf (r)\right] \\ \nonumber && \hspace{5mm}
+ \frac{1}{2}
\left[\frac{d}{dr}H(r) \tilde f_\phi(r)- H(r)\frac{d}{dr}\tilde f_\phi(r)
\right] +\frac{1}{\xi}\frac{d}{dr} \tilde f_\calf(r)
\eea
as a consistency condition for $\calr$. It can be solved readily
\bea 
\calr(r)&=& \frac{\xi}{2} H(r)\tilde f_\phi(r)-\tilde f_\calf(r)
 \\ \nonumber
&&\hspace{5mm} 
 +\frac{1}{2 r^2} \int_0^r dr'r'^2
\left[H'(r')\tilde f_\phi(r')-H(r')\tilde f'_\phi(r')
+ \frac{2}{\xi} \tilde f'_\calf(r')\right]
\pkt \eea
This fixes the right hand side of equation (\ref{dglFb}) for
$f_g(r)$ which is one of the basic ones for the new amplitudes.
The equations for the other amplitudes become
\bea
 \label{dgl3t}
&&-\tilde f_\phi'' 
-\frac{2}{r} \tilde f_\phi' + \frac{\llp}{ r^2} \tilde f_\phi +
\left\{ m_H^2 + 
\frac{\lambda}{g^2}\left[H^2(r)-\frac{3}{4}\epsilon H(r)\right]
\right\}
\tilde f_\phi 
\\ \nonumber && = \omega^2 \tilde f_\phi
-\frac{1}{2}H(r)\tilde f_\calf
+\frac{H(r)}{4 r^2} \int_0^r dr'r'^2
\left[H'(r')\tilde f_\phi(r')-H(r')\tilde f'_\phi(r')
+ \frac{2}{\xi} \tilde f'_\calf(r')\right]
\\
\label{dglFt}
&& - \tilde f''_\calf-\frac{2}{r}  \tilde f'_\calf 
+\frac{\llp}{r^2} \tilde f_\calf
+\xi \frac{H^2(r)}{4} \tilde f_\calf
\\ \nonumber
&=& \omega^2 \left\{-\xi \frac{H(r)}{2} \tilde f_\phi -
 \frac{1}{2 r^2} \int_0^r dr'r'^2
\left[H'(r')\tilde f_\phi(r')-H(r')\tilde f'_\phi(r')
+ \frac{2}{\xi} \tilde f'_\calf(r')\right] \right\}  
\pkt \eea  
Obviously, we have not succeeded in separating the system. 
However, in this form the gauge and gauge-fixing modes 
are easy to identify.
We see that with the choice  $\tilde f_\phi=0$ 
and $\tilde f_\calf=0$ the function $\calr(r)$ vanishes and 
the differential equation for $f_g$ becomes the Faddeev-Popov
equation again, with a corresponding energy spectrum.
Likewise, the combination $f_\calf=\tilde f_\calf 
+\omega^2 f_g$ still satisfies (\ref{dglF}) and has a Faddeev-Popov
eigenvalue spectrum as well. However, we do not find another linearly
independent combination of
amplitudes involving the  amplitude $\tilde f_\phi$ 
that would satisfy a differential
equation independent of $\xi$. So that part of the energy spectrum
that is not compensated by the Faddeev-Popov contributions
apparently still depends  on the choice of $\xi$.

Matters are different, however, if we evaluate the effective action.
This can be done using the fluctuation modes at
$\omega=0$, using a general theorem on fluctuation
determinants \cite{CoAS}, generalized to coupled systems, that has been
used, e.g., for computing the fluctuation corrections to bubble
nucleation \cite{Baa:1995a}.
It is based on the equation \footnote{For a short proof along the lines
of \cite{CoAS} see \cite{JBDiv}.}
\be \label{Flucdef} 
{\cal J}(\nu) \equiv \frac{\det (\bfm +\nu^2)}{\det (\bfm_0+\nu^2)}
= \lim_{r \to \infty} \frac {\det {\bf f}(\nu,r)}{\det
{\bf f}_0(\nu,r)} 
\pkt \ee
Here $\bfm$ is the partial wave fluctuation operator as defined
previously, and the matrix ${\bf f}(\nu,r)$ is an 
$(n\times n)$ matrix formed by a fundamental system of
$n$ linearly independent $n$-tuples of solutions for a given $\nu$,
regular at $r=0$.
The operator $\bfm_0$ and the solutions ${\bf f}_0$ refer
to a trivial background field configuration, in the present case
to the symmetric vacuum state characterized by $H(r)\equiv 0$.
It is understood, that both systems ${\bf f}$ and ${\bf f}_0$
are started, at $r=0$ with identical initial conditions.
Finally, the desired fluctuation determinant is given by
$\calj \equiv \calj(0)$. 

If we apply the theorem we only need the coupled system
of differential equations for $\omega=i\nu=0$, and then it
decouples in a triangular way. The right hand side
of the equation for $\tilde f_\calf$ vanishes entirely,  the r.h.s.
of the differential equation for $\tilde f_\phi$ only depends on
$\tilde f_\calf$, while both $\tilde f_\phi$ and
$\tilde f_\calf$ appear on the r.h.s. of the equation for 
$f_g$. Furthermore, for $\tilde f_\calf=0$, the differential
equation for $\tilde f_\phi$ becomes independent of $\xi$.
We can choose the following set of linearly independent
solutions:

(i) a gauge mode solution $f_n^g$ with $\tilde f_\calf^g \equiv 0$ and
 $\tilde f_\phi^g \equiv 0$, for which   $f_g^g$ evolves in the
same way as a pure Faddeev-Popov mode;

(ii) a `physical' solution $f_n^\phi$
 with $\tilde f_\calf^\phi \equiv 0$;
then  $\tilde f_\phi^ \phi$ evolves independently; it appears
on the right hand side of the differential equation for
$f_g^\phi $, which can be obtained by using the Green function
of the homogenous equation, and finally

(iii) a gauge-fixing mode solution $f_n^\calf$,
 where  $\tilde f^\calf_\calf$
is different from zero. For $\nu=0$ the r.h.s. of
(\ref{dglFt}) vanishes and $\tilde f_\calf$
evolves like a Faddeev-Popov mode.
 Both other amplitudes are different from zero in this case.
Note that the second type of solution is determined only
modulo an arbitrary multiple of the first one,
and the third one only modulo arbitrary multiples of
both other ones. This does not affect the determinant
$\det {\bf f}(0,r) $, however.
 
The structure of the matrix
${\bf f}(0,r) $ now is triangular and its determinant is obtained from
the diagonal elements as
\be
\det{\bf f}(0,r)  = f_g^g(0,r)\tilde f_\calf^\calf(0,r)
\tilde f_\phi^\phi(0,r) = f_{\rm FP}^2(0,r) \tilde f_\phi^\phi(0,r)
\pkt \ee
The same structure holds for the free solutions which have to
be started at $r=0$ with identical initial conditions, i.e.
with the same coefficients of the lowest powers of $r$, as determined
by the centrifugal barriers. We have considered the behavior
 at $r=0$ in detail and have verified that an
appropriate choice  is possible.

The effective action is obtained by adding the logarithms of
the various fluctuation determinants for all independent
systems, and for all partial waves. The only $\xi$ dependence
occurs in the gauge and gauge-fixing modes of the coupled system,
and for the two Faddeev-Popov modes. As these compensate each other
the total effective action becomes independent of $\xi$.

For the practical computation this means that for the
coupled system we just have to solve the integro-differential
equation for $\tilde f_\phi$ with $\tilde f_\calf=0$, i.e.,
\bea
 \label{dgl33t}
&&-\tilde f_\phi'' 
-\frac{2}{r} \tilde f_\phi' + \frac{\llp}{ r^2} \tilde f_\phi +
\left\{ m_H^2 + 
\frac{\lambda}{g^2}\left[H^2(r)-\frac{3}{4}\epsilon H(r)\right]
\right\}
\tilde f_\phi 
\\ \nonumber && =
\frac{H(r)}{4 r^2} \int_0^r dr'r'^2
\left[H'(r')\tilde f_\phi(r')-H(r')\tilde f'_\phi(r')\right]
\pkt\eea

From this derivation and discussion it is clear that the gauge
independence only holds for the effective action, and not
for other physical quantities. The nondiagonal parts of
the mode solutions still depend on $\xi$, so other expectation
values are affected by the gauge parameter $\xi$.


\section {Conclusion} 

We have given general arguments, based on the fluctuation operator
and the mode expansion, for the gauge independence of the
one-loop effective action, computed for a background field
which solves the classical equation of motion.
There are various cases for which the one-loop
effective action, and its gauge independence, are 
of interest. It appears in particular in 
the corrections to quantum or thermal tunneling rates
obtained in the semiclassical approximation.
 For the case of bubble nucleation the analysis of
section 3 fully applies. We have verified 
in the partial wave mode equations \cite{Baa:1995a}, that the
system of these differential equations can be cast, at
zero frequency, into a
triangular form. A theorem on
fluctuation determinants relates the fluctuation
determinant to the asymptotic behavior as $r \to \infty$, of a
linearly independent system of solutions regular
at $r=0$. The matrix formed by the $n$ linearly independent
$n$-tuples of  solutions can be cast into a triangular 
form as well, two of the diagonal elements evolve like the
Faddeev-Popov modes. Their contribution to the
logarithm of the fluctuation determinant is
cancelled by the one of the Faddeev-Popov modes.
The remaining diagonal elements are independent of $\xi$.
The final conclusion is that the {\em exact} one-loop
correction to the nucleation rate is gauge independent.
This  goes beyond the results of Ref. \cite{Metaxas:1996},
where a similar statement was derived for the leading orders
in the gauge coupling, using the gradient expansion. This latter
publication is, in part, complementary to our work: 
we have not considered
the divergent parts and renormalization. Within our method
\cite{Baa:1993b,Baa:1995a} the divergent
parts can be separated analytically from the computation
of the determinants and take the form of ordinary Feynman graphs.
So the $\xi$-independence of the renormalized
leading order contributions, as established
in \cite{Metaxas:1996}, closes our argument.

We should like to add a comment on the use of 
partial resummations.
Indeed the high temperature effective potential
(\ref{htpot2}), from which the 
classical solution is computed, already contains one-loop effects.
This introduces some double-counting that has to be compensated for
\cite{Baacke:1995b}. If the resummation includes the coupled 
gauge-Higgs sector, the $\xi$ dependence is there from the outset
and can  disappear only if all higher-loop orders 
are summed up. One therefore has to make sure that
the high temperature resummation,
that in an essential way
determines the structure of the phase transition, takes into
account transverse gauge loops and  the
isoscalar Higgs loop only. 
Then this modification of the  ``classical'' Higgs potential 
does not interfere with our analysis.
 
It is not clear how far the conclusions obtained for the special
case considered here can be generalized to different 
gauge theories, and to different physical systems.
For the sphaleron \cite{Baacke:1994a}, 
and also for topologically nontrivial
solutions in other models, as the instanton of
the abelian Higgs model in $(1+1)$ dimensions
\cite{Baacke:1995b}, the application of the
determinant theorem meets difficulties \cite{BaaKi_unp}:
The contribution of the $s$-wave diverges and the compensation
of this divergence by the sum over the higher partial waves
requires a suitable regularization. It would be worthwhile to
pursue this issue. 
 
Another system which should be investigated are
the quantum fluctuations for the $SU(2)$ Higgs model
in nonequilibrium dynamics \cite{Baacke:1997b}.
Here the applicability is certainly limited by the
inclusion of the quantum back reaction on the background
field. This back reaction changes the classical equation of motion,
while analogous changes of the quantum mode equations depend
on the resummation.

\section*{Acknowledgments}
The authors thank H. de Vega and the other members of
the LPTHE at the Universit\'e 
Pierre et Marie Curie for the warm hospitality extended to them.



\begin{thebibliography}{99}



\bibitem{Enq:1992}
K. Enqvist, J. Ignatius, K. Kajantie, and K. Rummukainen, Phys. Rev. D
{\bf 45}, 3415 (1992) .

\bibitem{Baa:1993b}
J. Baacke and V. G. Kiselev, Phys. Rev. D {\bf 48}, 5648 (1993) .

\bibitem{Baa:1995a}
 J. Baacke, Phys. Rev. D {\bf 52}, 6760 (1995) .

\bibitem{Kripf:1995}
J. Kripfganz, A. Laser, and M. G. Schmidt, Nucl. Phys. B {\bf 433}, 467
(1995) .

\bibitem{Baa:1996}
J. Baacke and A. S\"urig, Phys. Rev. D {\bf 52}, 4499 (1996) .

\bibitem{Sur:1998}
A. S\"urig, Phys. Rev. D {\bf 57}, 5049 (1998) .


\bibitem{Carson:1990a}
L. Carson, X. Li, and L. McLerran, Phys. Rev. D {\bf 42}, 2127 (1990) .

\bibitem{Carson:1990b}
L. Carson and L. McLerran, Phys. Rev. D {\bf 41}, 647 (1990) .

\bibitem{Baacke:1994a}
J. Baacke and S. Junker, Phys. Rev. D {\bf 49}, 2055 (1994);
{\em ibid.} {\bf 50}, 4227 (1994) .

\bibitem{Diakonov:1994}
D. Diakonov et al., Phys. Lett. B {\bf 336}, 457 (1994) .


\bibitem{Baacke:1995b}
J. Baacke and T. Daiber, Phys. Rev. D {\bf 51}, 795 (1995) .




\bibitem{Nielsen:1975}
N. K. Nielsen, Nucl. Phys. B{\bf 101}, 173 (1975) .

\bibitem{Aitchison:1984}
I. J. R. Aitchison and C. M. Fraser, Ann. Phys. {\bf 156}, 1 (1984) .

\bibitem{Kobes:1991}
R. Kobes, G. Kunstatter, and A. Rebhan,
Nucl. Phys. {\bf B355}, 1 (1991).

\bibitem{Contreras:1997}
C. Contreras and L. Vergara, Phys. Rev. D {\bf 55}, 5241 (1997) .

\bibitem{Metaxas:1996}
D. Metaxas, E. J. Weinberg, Phys. Rev. D {\bf 53}, 836 (1996) .


\bibitem{Baacke:1997b}
J. Baacke, K. Heitmann, and C. P\"atzold, 
Phys.  Rev. D {\bf 55}, 7815 (1997).

\bibitem{Boyanovsky:1996}
D. Boyanovsky, D. Brahm, R. Holman, and D. S. Lee, Phys. Rev. D {\bf 54},
1763 (1996) .

\bibitem{Boyanovsky:1998}
D. Boyanovsky, Will Loinaz, and R. S. Willey, Phys. Rev. D {\bf 57},
100 (1998) .



\bibitem{DLHLL} M. Dine, R. G. Leigh, P. Huet, A. Linde
and D. Linde, Phys. Rev. {\bf D46}, 550 (1992) .

\bibitem{HeKriSch}M. Hellmund, J. Kripfganz and
M. G. Schmidt, Phys. Rev. {\bf D50}, 7650 (1994) .

\bibitem{KriLaSch} J. Kripfganz, A. Laser and M. G. Schmidt,
Nucl. Phys. {\bf B433}, 467 (1995) .

\bibitem{La} J. S. Langer,  Ann. Phys. (N. Y.)
{\bf 41}, 108 (1967); {\it ibid.} {\bf 54}, 258 (1969) .

\bibitem{Co}S. Coleman, Phys. Rev. {\bf D15}, 2929 (1977) .

\bibitem{CaCo}G. Callan and S. Coleman, 
Phys. Rev. {\bf D16}, 1762 (1977) .

\bibitem{Af}  I. Affleck, Phys. Rev. Lett. {\bf 46},
388 (1981) .  

\bibitem{Li} A. D. Linde,  Nucl. Phys. {\bf B216}, 421 (1983) .

\bibitem{CoAS} see e.g. S. Coleman, {\it The Uses of Instantons},
in {\it The Aspects of Symmetry}, Cambridge University Press (1985) .

\bibitem{JBDiv} J. Baacke, 
{\it Fluctuation corrections to bubble nucleation}, 
in {\it Hot Hadronic Matter}, J. Letessier, H. H. Gutbrod, and
J. Rafelski, Eds., Plenum Press, New York 1995.

\bibitem{BaaKi_unp}
J. Baacke and V. G. Kiselev, unpublished work.

 




\end{thebibliography}
\end{document}